%% file: eprint.tex
\documentclass[12pt]{article}
\usepackage{graphicx}
\usepackage{amsmath}
\usepackage{amssymb}

\newcommand\pubnumber{CIPANP2018-Callahan}
\newcommand\pubdate{\today}

\def\iub{Department of Physics\\
Indiana University Bloomington, IN 47405}
\def\support{\footnote{Work supported by the National Science Foundation}}

\def\Title#1{\begin{center} {\Large #1 } \end{center}}
\def\Author#1{\begin{center}{ \sc #1} \end{center}}
\def\Address#1{\begin{center}{ \it #1} \end{center}}

\newcommand\pubblock{\rightline{\begin{tabular}{l} \pubnumber\\
         \pubdate  \end{tabular}}}
\newenvironment{Abstract}{\begin{quotation}  }{\end{quotation}}
\newenvironment{Presented}{\begin{quotation} \begin{center} 
             PRESENTED AT\end{center}\bigskip 
      \begin{center}\begin{large}}{\end{large}\end{center} \end{quotation}}

\input econfmacros.tex
\begin{document}
\begin{titlepage}
\pubblock

\vfill
\Title{Measurement of the neutron lifetime using a magneto-gravitational trap}
\vfill
\Author{Nathan Callahan\support\\for the UCN\(\tau\) Collaboration}
\Address{\iub}
\vfill
\begin{Abstract}
Precision measurements of the free neutron lifetime \(\tau_n\), when combined with measurements of the axial vector form factor, can be used to test unitarity of the CKM matrix. Non-unitarity is a signal for physics Beyond the Standard Model (BSM). Sensitivity to BSM physics requires measurements of \(\tau_n\) to a precision of 0.1~s. However, the two dominant techniques to measure \(\tau_n\) (colloquially beam and bottle measurements) disagree by nearly 10~s. UCN\(\tau\) is a neutron lifetime experiment using a magneto-gravitational trap and an \textit{in-situ} neutron detector. Neutrons in this trap are not susceptible to loss on material walls as in previous bottle measurements. Additionally, the \textit{in-situ} detector allows spectral monitoring of the trapped Ultracold Neutrons. In this talk, I will present our most recent result - \(\tau_n=877.7\pm0.7_\text{(stat.)}+0.4/-0.2_\text{(sys.)}\)~s. I will also present Monte Carlo simulations of systematic effects in the experiment including heating and spectral cleaning.
\end{Abstract}
\vfill
\begin{Presented}
Conference on the Intersections of Particle and Nuclear Physics\\
Palm Springs, CA May 29 - June 3 2018
\end{Presented}
\vfill
\end{titlepage}
\def\thefootnote{\fnsymbol{footnote}}
\setcounter{footnote}{0}

\section{Introduction}
Neutrons undergo free \(\beta\) decay via the weak force. The mean lifetime of neutron decay is approximately 15~minutes. The neutron lifetime (\(\tau_n\)) is an important parameter to models of Big Bang Nucleosynthesis (where it can predict the primordial helium abundance \(Y_p\)) \cite{Dubbers:2011ns}. Additionally, \(\tau_n\), when combined with other \(\beta\) decay observables, can serve as a measurement of \(V_{ud}\) (the CKM matrix element) \cite{Cirigliano:2013xha} complimentary to the much more precise determinations using superallowed nuclear \(\beta\) decay \cite{Towner:2010zz}. There are two major ways to measure the neutron lifetime: measuring surviving neutrons in a trap, or measuring decay products in a neutron beamline. Within the last decade, a significant discrepancy (approximately 4\(\sigma\)) has emerged between the two methods, with beams measuring 888.0\(\pm\)2~s and traps measuring 878.1\(\pm\)0.5~s. A shift of 10~s changes the prediction of the primordial helium abundance by approximately the uncertainty of observational values \cite{Pitrou:2018cgg}.

The two experimental methods to determine \(\tau_n\) are susceptible to different systematic effects: for trap measurements non-\(\beta\) decay losses and for beam measurements characterization of detector efficiency. The Particle Data Group average currently stands at 880.2\(\pm\)1~s \cite{Tanabashi:2018oca}. For trap measurements, significant extrapolations have to be made if the traps lose neutrons during storage. Historically, this has been as low as 10~s and upwards of 100~s. The measurements which produce the average are reproduced in Table~\ref{tab:pdg}, along with the approximate extrapolations and systematic errors reported (see ref. \cite{Tanabashi:2018oca} for citations).

These large extrapolations can be avoided if a different trap is used. Instead of trapping ultracold neutrons using material potentials, neutrons can be trapped by magnetic fields. Neutrons experience a potential of \(\sim60\)~neV/T and lose \(\sim\)1~neV/cm in earth's gravitational field. Ultracold neutrons can adiabatically follow the local magnetic field and so can see a potential proportional to the field strength as long as they maintain their polarization. Absent depolarization or scattering on residual gas, neutrons can be stored in such a trap without losses other than \(\beta\) decay.

\begin{table}[htb]
\begin{center}
    \begin{tabular}{lllll}
    Author & \(\sigma_\text{stat.}\) [s] & \(\Delta \tau_\text{sys.}\) [s] & Extrapolation [s] & Method \\ \hline
    Arzumanov 2015 & 0.64 & 3.6 & 40-280 & Bottle \\
    Steyerl 2012 & 1.4 & \(\sim\)7 & \(>\)200~s & Bottle \\
    Pichlmaier 2010 & 1.3 & 1 & 110-300 & Bottle \\
    Serebrov 2005 & 0.7 & 0.4 & 10-20 & Bottle \\
    Yue 2013 & 1.2 & 1 & 2-15 & Beam \\
    Byrne 1996 & 3 & 5.9 & - & Beam \\
    \end{tabular}
    \caption{A selection of neutron Lifetime experiments with systematic corrections and approximate extrapolation scales}
    \label{tab:pdg}
\end{center}
\end{table}

\section{2016 Run Campaign}
\begin{figure}[htb]
    \centering
	\includegraphics[width=0.75\textwidth]{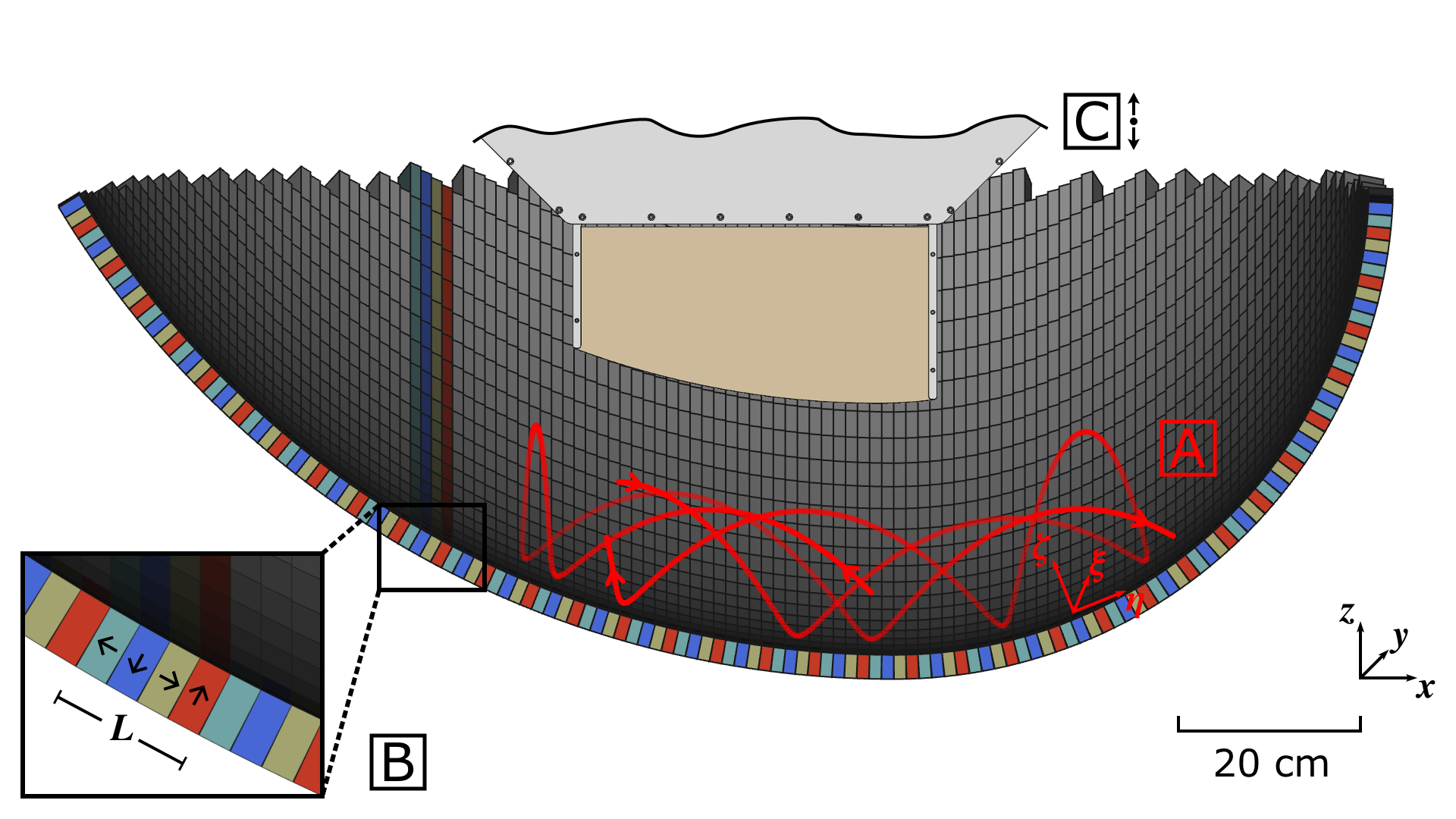}
	\caption{The UCN\(\tau\) apparatus. \textbf{A} UCN trajectory inside the trap which enters and leaves the page as indicated. \textbf{B} Arrows indicating the magnetization direction for the Halbach Array. Colored strips show rows of identical magnetization. \textbf{C} The \textit{in-situ} detector which can be raised out of the trap and lowered to the bottom. The khaki color is the active area of \(^{10}\)B.}
	\label{fig:trajintrap}
\end{figure}

The UCN\(\tau\) collaboration has conducted a multi-year experiment to measure \(\tau_n\) using a magnetic trap. Ultracold neutrons (UCN), with energies \(<\)50~neV are trapped using a Halbach array of permament magnets. A rendering of the trap (along with a simulated UCN trajectory and a rendering of the \textit{in-situ} detector) is shown in Figure~\ref{fig:trajintrap}. Note that neutrons are confined below by the magnetic field and above by gravity.

The UCN\(\tau\) apparatus measures the neutron lifetime by counting surviving neutrons inside a magneto-gravitational trap after varying holding times. Neutrons are filled into the trap from below for 150~s from a source of ultracold neutrons. After filling, a cut out segment of the Halbach array is raised into the bottom of the trap to complete the Halbach array. During filling and for 50~s after, a neutron scatterer is placed into the trap at 38~cm from the bottom. This eliminates high-energy UCN that have sufficient energy to bounce over the walls of the trap. After cleaning, neutrons are stored for a short period (20~s) or a long period (\(>\)1000~s). After storage, a neutron detector is lowered into the trap and the population is measured \textit{in-situ}. The lifetime is measured by taking the ratio of the population remaining after the short and long holding times and the difference in holding times:
\begin{equation}
\tau_n=\frac{\Delta T}{\text{ln}(N_s/N_l)},
\end{equation}
where \(\Delta T\) is the difference in the holding times, \(N_s\) is the population after a short holding time, and \(N_l\) is the population after a long holding time. The lifetime can be measured in run pairs and averaged to obtain a central value.

This apparatus was used over several months in 2016 to measure several million UCN. The UCN\(\tau\) collaboration conducted a blinded analysis and measured the neutron lifetime to be 877.7\(\pm\)0.7~(stat)+0.4/-0.2~(sys)~s \cite{Pattie:2017vsj}. The true lifetime was unknown to the analyzers, and the analysis was published once it was unblinded. Subsequent years of data have accumulated more statistics and a statistical uncertainty of \(\sim\)0.3~s; analysis is still ongoing.

\section{Assesment of Systematic Effects}
The dominant systematic effects are those of depolarization (\(<\)0.07~s), phase space evolution (\(\pm\)0.1~s), heating of neutrons during storage (\(<\)0.24~s), and insufficient cleaning (\(<\)0.07~s). These systematic effects were measured using data-driven methods during the 2016 campaign. The effects of heating and insufficient cleaning were simulated in subsequent work. During the CIPANP meeting, the effects of heating and insufficient cleaning were approached using (at the time) unpublished simulation estimates.

The systematic effect of Depolarization occurs due to a small probability for neutrons to flip their polarization during storage. A neutron which depolarizes will be attracted to the trap walls and is lost. This contributes to the measured trap lifetime and must be accounted for. A small perpendicular holding field (on the order of 100~G) is provided inside the trap to help maintain polarization of trapped neutrons. Simulation work has shown that the expected dependence of the depolarization lifetime on the holding field strength is \(\propto 1/B_\text{hold}^2\) \cite{Steyerl:2016jvy}, where \(B_\text{hold}\) is the strength of the holding field. In the UCN\(\tau\) experiment, the trap lifetime was measured as a function of \(B_\text{hold}\). The trap lifetime as a function of field strength was then fit and the depolarization lifetime was extracted. An upper bound on the losses due to depolarization can then be obtained, yielding \(<0.07\)~s.

Phase space evolution is the process of the distribution of UCN in phase space shifting in time. Neutrons in the trap typically do not undergo diffuse reflections on the walls; as a consequence, the neutron cloud can have a long time constant to equilibrate in phase space. The distribution in phase space can affect the draining time of the neutrons during the counting period. If neutrons are drained more quickly after a long period of storage in the trap, then the effective storage time would be less. This is addressed by measuring the mean arrival time of neutrons in the short and long lifetime runs and using these measurements to calculate the neutron lifetime. By observing the uncertainty on the measurements of the mean arrival time, an upper bound for uncertainty due to phase space evolution can be obtained, yielding \(\pm0.1\)~s.

\section{Monte Carlo Simulations}
A simulation of the experiment can be conducted in order to estimate shifts in the trap lifetime due to insufficient cleaning or heating. Individual neutrons were tracked inside the magnetic trap. The magnetic field was evaluated using a field expansion, which reproduces the exponentially decaying field from a Halbach array, as well as periodic ripples in the strength \cite{Walstrom}. The trapdoor was not simulated; however neutrons were created at the bottom of the trap to imitate their entry during the loading process. The distribution in phase space was modeled using a spectral distribution in energy and an initial angular distribution. The detector cross section was taken from CAD models and the surface was simulated as a multilayer quantum 1-D system. The cleaner was simulated inside the trap as well. Neutrons were born inside the trap, tracked during the cleaning and storage process, and drained into the detector.

The simulation has 5 free parameters: two controlling the spectral distribution, one controlling the forward-directedness during UCN creation, and two parameters describing the detector surface. The arrival time histogram of UCN during the counting period was compared to data taken in 2016. The 5 parameters were optimized using \(\chi^2\) minimization of the distance between the two histograms. This optimized model was then used to simulate an independent dataset with a different detector insertion profile.

One source of heating is microphonic vibrations of the trap. If the Halbach array vibrates in space, a neutron impinging on the magnetic field can gain or lose a small amount of energy. Over time, a neutron can undergo a random walk in energy. If a neutron gains sufficient energy, it can escape the trap. This loss mechanism will distort the measured lifetime and must be accounted for. Similarly, if high-energy neutrons remain inside the trap after cleaning, then they can escape as well.

The vibrational profile of the trap was measured using an accelerometer. This vibrational data was placed into the simulation to study the effects of heating. The vibrational amplitude was approximately 1~\(\mu\)m in amplitude.

Simulations of cleaning and nominal vibrational amplitudes show a shift in the trap lifetime of \(0.03\)~s due to insufficient cleaning and heating. Increasing the amplitude of vibrations to 40~\(\mu\)m showed a shift in the trap lifetime of \(0.15\)~s due to heating. The simulations show that the UCN\(\tau\) apparatus is nominally safe from microphonic heating and effective at cleaning.

\section{Conclusion and Implications}
The UCN\(\tau\) collaboration has measured the neutron lifetime to be 877.7\(\pm\)0.7~(stat)+0.4/-0.2~(sys)~s, which is consistent with previous bottle measurements. This result maintains the tension in the neutron lifetime field. Currently, the implications for Big Bang Nucleosynthesis are not significant; however, in the future more precise observational measurements may be able to differentiate between the beam and bottle measurements. The UCN\(\tau\) result, when combined with modern measurements of \(\lambda=g_A/g_V\) (the zero-momentum axial vector form factor), form a consistent picture of the Standard Model with superallowed \(\beta\) decays \cite{Brown:2017mhw}. UCN\(\tau\) currently has the data to perform a 0.3~s statistical measurement of \(\tau_n\), further improving its precision.

%
%\Acknowledgements
%I am grateful to Don Alfonso d'Alba for certain services essential to 
%this investigation.

\end{document}

%% file: econfmacros.tex
\def\beq{\begin{equation}}
\def\eeq#1{\label{#1}\end{equation}}
\def\eeqn{\end{equation}}

\def\beqa{\begin{eqnarray}}
\def\eeqa#1{\label{#1}\end{eqnarray}}
\def\eeqan{\end{eqnarray}}

\let\bar=\overbar

\def\Dslash{\not{\hbox{\kern-4pt $D$}}}
\def\dslash{\not{\hbox{\kern-2pt $\del$}}}

\def\msb{{\bar{\ssstyle M \kern -1pt S}}}

%% file: eprint.bbl
\begin{thebibliography}{99}

%\cite{Pattie:2017vsj}
\bibitem{Pattie:2017vsj} 
  R.~W.~Pattie, Jr. {\it et al.},
  %``Measurement of the neutron lifetime using a magneto-gravitational trap and in situ detection,''
  Science {\bf 360}, no. 6389, 627 (2018)
  doi:10.1126/science.aan8895
  [arXiv:1707.01817 [nucl-ex]].
  %%CITATION = doi:10.1126/science.aan8895;%%
  %27 citations counted in INSPIRE as of 01 Oct 2018

%\cite{Dubbers:2011ns}
\bibitem{Dubbers:2011ns} 
  D.~Dubbers and M.~G.~Schmidt,
  %``The Neutron and Its Role in Cosmology and Particle Physics,''
  Rev.\ Mod.\ Phys.\  {\bf 83}, 1111 (2011)
  doi:10.1103/RevModPhys.83.1111
  [arXiv:1105.3694 [hep-ph]].
  %%CITATION = doi:10.1103/RevModPhys.83.1111;%%
  %113 citations counted in INSPIRE as of 01 Oct 2018

%\cite{Cirigliano:2013xha}
\bibitem{Cirigliano:2013xha} 
  V.~Cirigliano, S.~Gardner and B.~Holstein,
  %``Beta Decays and Non-Standard Interactions in the LHC Era,''
  Prog.\ Part.\ Nucl.\ Phys.\  {\bf 71}, 93 (2013)
  doi:10.1016/j.ppnp.2013.03.005
  [arXiv:1303.6953 [hep-ph]].
  %%CITATION = doi:10.1016/j.ppnp.2013.03.005;%%
  %79 citations counted in INSPIRE as of 01 Oct 2018
  
%\cite{Towner:2010zz}
\bibitem{Towner:2010zz} 
  I.~S.~Towner and J.~C.~Hardy,
  %``The evaluation of V(ud) and its impact on the unitarity of the Cabibbo-Kobayashi-Maskawa quark-mixing matrix,''
  Rept.\ Prog.\ Phys.\  {\bf 73}, 046301 (2010).
  doi:10.1088/0034-4885/73/4/046301
  %%CITATION = doi:10.1088/0034-4885/73/4/046301;%%
  %107 citations counted in INSPIRE as of 01 Oct 2018
  
%\cite{Pitrou:2018cgg}
\bibitem{Pitrou:2018cgg} 
  C.~Pitrou, A.~Coc, J.~P.~Uzan and E.~Vangioni,
  %``Precision big bang nucleosynthesis with improved Helium-4 predictions,''
  Phys.\ Rept.\  {\bf 04}, 005 (2018)
  doi:10.1016/j.physrep.2018.04.005
  [arXiv:1801.08023 [astro-ph.CO]].
  %%CITATION = doi:10.1016/j.physrep.2018.04.005;%%
  %13 citations counted in INSPIRE as of 01 Oct 2018
  
%\cite{Tanabashi:2018oca}
\bibitem{Tanabashi:2018oca} 
  M.~Tanabashi {\it et al.} [Particle Data Group],
  %``Review of Particle Physics,''
  Phys.\ Rev.\ D {\bf 98}, no. 3, 030001 (2018).
  doi:10.1103/PhysRevD.98.030001
  %%CITATION = doi:10.1103/PhysRevD.98.030001;%%
  %219 citations counted in INSPIRE as of 01 Oct 2018
  
%\cite{Steyerl:2016jvy}
\bibitem{Steyerl:2016jvy} 
  A.~Steyerl, K.~K.~H.~Leung, C.~Kaufman, G.~Müller and S.~S.~Malik,
  %``Spin flip loss in magnetic confinement of ultracold neutrons for neutron lifetime experiments,''
  Phys.\ Rev.\ C {\bf 95}, no. 3, 035502 (2017)
  doi:10.1103/PhysRevC.95.035502
  [arXiv:1611.05037 [physics.ins-det]].
  %%CITATION = doi:10.1103/PhysRevC.95.035502;%%
  %1 citations counted in INSPIRE as of 01 Oct 2018
  
\bibitem{Walstrom}
    P.L. Walstrom and J.D. Bowman and S.I. Penttila and C. Morris and A. Saunders, NIM-A {\bf 599}, no. 1, 82 (2009) doi:10.1016/j.nima.2008.11.010
    
%\cite{Brown:2017mhw}
\bibitem{Brown:2017mhw} 
  M.~A.-P.~Brown {\it et al.} [UCNA Collaboration],
  %``New result for the neutron $\beta$-asymmetry parameter $A_0$ from UCNA,''
  Phys.\ Rev.\ C {\bf 97}, no. 3, 035505 (2018)
  doi:10.1103/PhysRevC.97.035505
  [arXiv:1712.00884 [nucl-ex]].
  %%CITATION = doi:10.1103/PhysRevC.97.035505;%%
  %14 citations counted in INSPIRE as of 01 Oct 2018




\end{thebibliography}
